\documentclass[11pt]{article}
\usepackage{graphicx}
\usepackage{epsfig}

\newcommand{\BABARPubYear}    {00}

\newcommand{\BABARProcNumber} {15}
\newcommand{\SLACPubNumber} {8709}

\def\lbabar{\mbox{{\large\sl B}\hspace{-0.4em} {\normalsize\sl A}\hspace{-0.03em}{\large\sl B}\hspace{-0.4em} {\normalsize\sl A\hspace{-0.02em}R}}}
\usepackage{relsize}
\def\babar{\mbox{\slshape B\kern-0.1em{\smaller A}\kern-0.1em
    B\kern-0.1em{\smaller A\kern-0.2em R}}}

\def\en         {\ensuremath{e^-}}      
\def\ep         {\ensuremath{e^+}}
  
\def\epem       {\ensuremath{e^+e^-}}

\def\qqbar {\ensuremath{q\overline q}}

\def\piz   {\ensuremath{\pi^0}}

\def\pim   {\ensuremath{\pi^-}}

\def\Kbar  {\ensuremath{\kern 0.2em\overline{\kern -0.2em K}}}

\def\Kp    {\ensuremath{K^+}}

\def\Kstarz  {\ensuremath{K^{*0}}}

\def\Kstarb   {\ensuremath{{\Kbar}\kern0.03em {\raise.18ex\hbox{$^*$}}}}

\def\Kzb   {\ensuremath{{\Kbar}\kern0.03em {\raise.1ex\hbox{$^0$}}}}
\def\KzKzb {\ensuremath{K^0 {\kern -0.16em \Kzb}}}

\def\Dbar  {\ensuremath{\kern 0.2em\overline{\kern -0.2em D}}}
\def\Dzb   {\ensuremath{{\Dbar}\kern0.03em {\raise.3ex\hbox{$^0$}}}}
\def\DzDzb {\ensuremath{D^0 {\kern -0.16em \Dzb}}}

\def\Dstarb   {\ensuremath{{\Dbar}\kern0.03em {\raise.18ex\hbox{$^*$}}}}

\def\Bz    {\ensuremath{B^0}}

\def\Bbar  {\ensuremath{\kern 0.18em\overline{\kern -0.18em B}}}
\def\Bzb   {\ensuremath{{\Bbar}\kern0.03em {\raise.3ex\hbox{$^0$}}}}
\def\Bu    {\ensuremath{B^+}}

\def\BB    {\ensuremath{B\Bbar}} 
\def\BzBzb {\ensuremath{B^0 {\kern -0.16em \Bzb}}}

\def\jpsi  {\ensuremath{{J\mskip -3mu/\mskip -2mu\psi\mskip 2mu}}} 
\def\psitwos {\ensuremath{\psi{(2S)}}}
\mathchardef\Upsilon="7107
\def\Y#1S{\ensuremath{\Upsilon{(#1S)}}}

\mathchardef\Delta="7101
\mathchardef\Xi="7104
\mathchardef\Lambda="7103
\mathchardef\Sigma="7106
\mathchardef\Omega="710A
\def\Deltabar   {\ensuremath{\kern 0.25em\overline{\kern -0.25em \Delta}}{}}
\def\Lbar {\ensuremath{\kern 0.2em\overline{\kern -0.2em\Lambda\kern 0.05em}\kern-0.05em}{}}
\def\Sigbar{\ensuremath{\kern 0.2em\overline{\kern -0.2em \Sigma}}{}}
\def\Xibar{\ensuremath{\kern 0.2em\overline{\kern -0.2em \Xi}}{}}
\def\Obar{\ensuremath{\kern 0.2em\overline{\kern -0.2em \Omega}}{}}
\def\Nbar{\ensuremath{\kern 0.2em\overline{\kern -0.2em N}}{}}

\def\BR{{\ensuremath{\cal B}}}

\def\btosgam   {\ensuremath{b \to s \gamma}}
\def\btodgam   {\ensuremath{b \to d \gamma}}

\def\mes        {\mbox{$m_{\rm ES}$}}

\def\ev   {\ensuremath{\rm \,e\kern -0.08em V}}
\def\kev  {\ensuremath{\rm \,ke\kern -0.08em V}} 
\def\mev  {\ensuremath{\rm \,Me\kern -0.08em V}} 
\def\gev  {\ensuremath{\rm \,Ge\kern -0.08em V}} 
\def\gevc {\ensuremath{{\rm \,Ge\kern -0.08em V\!/}c}} 
\def\tev  {\ensuremath{\rm \,Te\kern -0.08em V}}
\def\mevc {\ensuremath{{\rm \,Me\kern -0.08em V\!/}c}} 
\def\gevcc{\ensuremath{{\rm \,Ge\kern -0.08em V\!/}c^2}} 
\def\mevcc{\ensuremath{{\rm \,Me\kern -0.08em V\!/}c^2}}

\def\invfb   {\ensuremath{\mbox{\,fb}^{-1}}}
\def\mus  {\ensuremath{\rm \,\mus}}

\def\mus        {\ensuremath{\,\mu{\rm s}}}    
\def\gsim{{~\raise.15em\hbox{$>$}\kern-.85em
          \lower.35em\hbox{$\sim$}~}}
\def\lsim{{~\raise.15em\hbox{$<$}\kern-.85em
          \lower.35em\hbox{$\sim$}~}}

\def\to                 {\ensuremath{\rightarrow}}

\def\pep2{PEP-II}

\def\chic1{\ensuremath{\chi_{c1}}}
\def\chic2{\ensuremath{\chi_{c2}}}
\def\chic3{\ensuremath{\chi_{c3}}}

\newcommand{\eqref}[1]{Eq.~(\ref{eq:#1})}

\newcommand{\epjc}      [1]  {{Eur.\ Phys.\ Jour.\ C~{\bf #1}}}

\newcommand{\nim}       [1]  {{Nucl.\ Instr.\ and Methods~{\bf #1}}}
\newcommand{\nima}      [1]  {{Nucl.\ Instr.\ Methods {\bf A{\bf #1}}}}

\newcommand{\npb}       [1]  {{Nucl.\ Phys.\ {\bf B{\bf #1}}}}

\newcommand{\npbps}     [1]  {{Nucl.\ Phys.\ B Proc.\ Suppl.\ {\bf #1}}}

\newcommand{\prl}       [1]  {{Phys.\ Rev.\ Lett.\ {\bf #1}}} 
\newcommand{\pr}        [1]  {{Phys.\ Rev.\ {\bf #1}}}
\newcommand{\progtp}    [1]  {{Prog.\ Th.\ Phys.\ {\bf #1}}}
\newcommand{\rmp}       [1]  {{Rev.\ Mod.\ Phys.\ {\bf #1}}}  

\newcommand{\zpc}       [1]  {{Z.\ Phys.\ C~{\bf #1}}}

\newcommand{\mpl}     [1]  {{Mod.\ Phys.\ Lett.\ {\bf #1}}}

\def\jetset74   {\mbox{\tt Jetset \hspace{-0.5em}7.\hspace{-0.2em}4}}

\def\bkog    {\ensuremath {\Bz \to \Kstarz \gamma}}
\def\kpi    {\ensuremath {\Kstarz \to \Kp \pim}}
\def\brog    {\ensuremath {\Bz \to \rho \gamma}}

\def\mb      {\ensuremath {M_{ES}}}
\def\ebeam     {\ensuremath {E^{*}_{b}}}
\def\egcms     {\ensuremath {E^{*}_{\gamma}}}
\def\de        {\ensuremath {\Delta E^{*}}}
\def\mkpi      {\ensuremath {M_{\Kp \pim}}}
\def\mpl #1 #2 #3 {Mod.~Phys.~Lett.~{\bf#1},\ #2 (#3)}
\def\npb  #1 #2 #3 {Nucl.~Phys.~B~{\bf#1},\ #2 (#3)}
\def\plb  #1 #2 #3 {Phys.~Lett.~B~{\bf#1},\ #2 (#3)}
\def\pr   #1 #2 #3 {Phys.~Rep.~{\bf#1},\ #2 (#3)}
\def\prd  #1 #2 #3 {Phys.~Rev.~D~{\bf#1},\ #2 (#3)}
\def\prl  #1 #2 #3 {Phys.~Rev.~Lett.~{\bf#1},\ #2 (#3)}
\def\RMP  #1 #2 #3 {Rev.~Mod.~Phys.~{\bf#1},\ #2 (#3)}
\def\zpc  #1 #2 #3 {Z.~Phys.~C~{\bf#1},\ #2 (#3)}
\def\nim  #1 #2 #3 {Nucl.~Instrum.~Methods~{\bf#1},\ #2 (#3)}
\def\nima  #1 #2 #3 {Nucl.~Instrum.~Methods~A.{\bf#1},\ #2 (#3)}
\def\epjc #1 #2 #3 {Euro.~Phys.~Jour~{\bf#1},\ #2 (#3)}
\def\rmp #1 #2 #3 {Rev.~Mod.~Phys~{\bf#1},\ #2 (#3)}
\def\npbps #1 #2 #3 {Nucl.~Phys.~B.~proc.~suppl~{\bf#1},\ #2 (#3)}
\def\progtp #1 #2 #3 {Prog.~Theo.~Phys~{\bf#1},\ #2 (#3)}
\def\etal{{\it et al.}}

\def\brresult    {\ensuremath {\BR(\bkog)= (5.42 \pm 0.82(stat.) \pm 0.47(sys.)) \times 10^{-5}}}
\def\lumi    {\ensuremath {(8.6 \pm 0.3) \times 10^6}}

\long\def\inst#1{\par\nobreak\kern 4pt\nobreak
    {\it #1}\par\vskip 10pt plus 3pt minus 3pt}

\begin{document}
{\pagestyle{empty}

\begin{flushright}
SLAC-PUB-\SLACPubNumber \\
\babar-PROC-\BABARPubYear/\BABARProcNumber \\
August, 2000 \\
\end{flushright}

\par\vskip 3cm

\begin{center}
\Large \bf 
 First Results in Exclusive Radiative Penguin Decays at \babar
\end{center}
\bigskip

\begin{center}
\large 
Colin Jessop\\
(for the \lbabar\ Collaboration)\\
Stanford Linear Accelerator Center\\
Stanford, CA 94304, USA\\
E-mail: jessop@slac.stanford.edu
\end{center}
\bigskip \bigskip

\begin{center}
\large \bf Abstract
\end{center}
We present a preliminary measurement of the branching fraction of the exclusive
penguin decay $\bkog$ using \lumi\ $\BB$ decays 
\begin{eqnarray*}    
\brresult.
\end{eqnarray*}     
In addition we search for the related penguin decays with
a lepton pair in the final state,$B^+ \rightarrow K^+ l^+ l^-$,
$B^0 \rightarrow \Kstarz \ l^+ l^-$. We find no evidence for these decays
in $3.7 \pm\ 0.1 \times\ 10^6$ \BB\ decays and set preliminary 90 \% C.L upper
limits of
\begin{eqnarray*}    
\BR(B^+ \rightarrow K^+ e^+ e^-)  & <  & 12.5 \times 10^{-6}, \\
\BR(B^+ \rightarrow K^+ \mu^+ \mu^-) &  < & \phantom{2}8.3 \times 10^{-6}, \\
\BR(B^0 \rightarrow \Kstarz e^+ e^-) & < &  24.1  \times  10^{-6}, \\
\BR(B^0 \rightarrow \Kstarz \mu^+ \mu^-) & < & 24.5 \times 10^{-6}. \\
\end{eqnarray*} 

\vfill
\begin{center}
Contributed to the Proceedings of the 30$^{th}$ International 
Conference on High Energy Physics, \\
7/27/2000---8/2/2000, Osaka, Japan
\end{center}

\vspace{1.0cm}
\begin{center}
{\em Stanford Linear Accelerator Center, Stanford University, 
Stanford, CA 94309} \\ \vspace{0.1cm}\hrule\vspace{0.1cm}
Work supported in part by Department of Energy contract DE-AC03-76SF00515.
\end{center}

\setlength\columnsep{0.20truein}
\twocolumn
\def\sloppy{\tolerance=100000\hfuzz=\maxdimen\vfuzz=\maxdimen}
\sloppy
\vbadness=12000
\hbadness=12000
\flushbottom
\def\figurebox#1#2#3{%
  	\def\arg{#3}%
  	\ifx\arg\empty
  	{\hfill\vbox{\hsize#2\hrule\hbox to #2{\vrule\hfill\vbox to #1{\hsize#2\vfill}\vrule}\hrule}\hfill}%
  	\else
   	{\hfill\epsfbox{#3}\hfill}%
  	\fi}

\section{Introduction}
In the Standard Model the exclusive decay $\Bz \to \Kstarz \gamma$ proceeds 
by the $\btosgam$ loop ``penguin'' diagram.
Precise measurements of decay modes involving these transitions and modes
with the related $\btodgam$  transition such as $\brog$ will allow
measurements of the top quark couplings $V_{ts}$ and $V_{td}$.
The strength of these transitions may also be  enhanced by the
presence of non-Standard Model contributions~\cite{joanne}. 
  In the first year of running the $\babar$ experiment has  accumulated
a dataset comparable to the world's largest to date, and  
this will increase by an order of magnitude over the next few years.
A comprehensive program to study these decays is now underway.
The first step in this program is the preliminary measurement of the 
branching fraction of the exclusive decay mode $\bkog$ using the leading decay mode, \kpi\ . Here \Kstarz\ refers to the $K^{*0}(896)$ resonance, and charge 
conjugate channels are assumed throughout. We also present a search for 
the rarer and as yet unobserved exclusive penguin decays
$B\to K\ell^+\ell^-$ and $B\to K^*\ell^+\ell^-$,
where $\ell$ is either an electron or muon.
 
The data were collected with the \babar\ detector
 at the \pep2\ asymmetric \epem\ storage ring.
The results presented in this paper are based upon an
integrated luminosity of 7.5 $\invfb$ of  data corresponding to
\lumi\ \BB\ meson pairs recorded at the $\Y4S$ energy (``on-resonance'') and
$1.1 \invfb$ below the $\Y4S$ energy (``off-resonance''). 
We compute quantities
in both the laboratory frame and the rest frame of the \Y4S\ . Quantities
computed in the rest frame are denoted by an asterisk; eg. $\ebeam$ is the
energy of the \ep\ and \en\ beams which are symmetric in the \Y4S\ rest frame.

\section{Measurement of $\BR(\bkog)$}
We begin the selection by requiring a good photon candidate in
the calorimeter with  an energy $2.20 \gev < \egcms < 2.85 \gev$. We veto photons from $\piz$'s. We next reconstruct the $\Kstarz$ from $\Kp$ and $\pim$
candidates. We consider all pairs of
tracks in the event. A  track is identified as a kaon by the 
ring imaging Cherenkov detector (DIRC) and we require $ 806 \mev < \mkpi\ 
< 986 \mev$. The \Bz\ candidates are reconstructed from the \Kstarz\ and $\gamma$
candidates. There are  backgrounds from continuum \qqbar\ production with the
high energy photon originating  from  initial state radiation or from a
$\piz$ or $\eta$. We exploit event topology differences between signal
and background to reduce the continuum contribution~\cite{ichepbkg}.
In the rest frame of the \Y4S\ the \BB\ pairs are
produced approximately at rest and therefore decay isotropically while the
\qqbar\ pair recoil against each other in
a jet-like topology. 

Since the \Bz\ mesons are produced via $\epem \to \Y4S \to \BB$ the energy of
the \Bz\ is given precisely by the beam energy, \ebeam\ . We 
reconstruct the \Bz\ candidate substituting  \ebeam\ for the
measured energy of the candidate daughters.  We define the difference
of the  beam energy and energy of the \Bz\ daughters, 
$\de = \ebeam - E^{*}_{K^*} - \egcms$ and require 
$ -200 \mev < \de < 100 \mev $. 
The \Bz\ mass is given by,
$\mb= \sqrt{E^{*2}_{beam}+|p_{B}^{*2}|}$, where $|p^{*}_{B}|$ is the momentum
of the \Bz\ candidate calculated using the measured momenta of the charged
daughters and the energy of the photon. Figure~\ref{fig:datamb} shows
the \mb\ of the candidates. The background is determined empiracally by fitting the ARGUS function~\cite{argusf}
to off-resonance data.
We find a signal of $48.4 \pm 7.3$ events with the error coming from 
the statistical error of the fit.  

\begin{figure}
\epsfxsize180pt
\figurebox{180pt}{160pt}{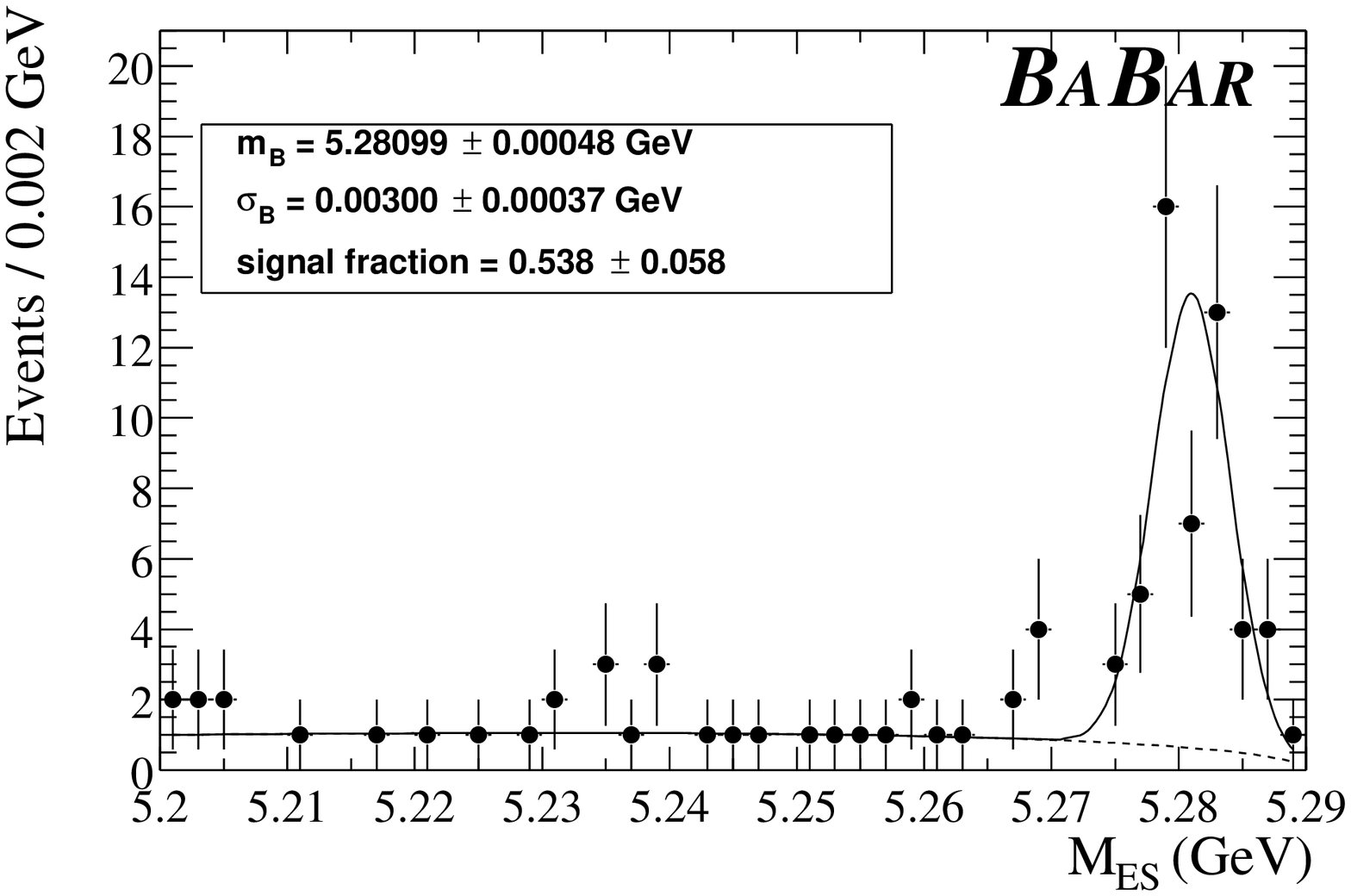}
\caption{The \mb\ projection  for \bkog\, \kpi\ candidates from \lumi\ 
\BB\ decays.}
\label{fig:datamb}
\end{figure}

\begin{table}[hbtb]
\caption{The fractional systematic uncertainties in the measurement of
\BR(\bkog).}
\begin{center}
\begin{tabular}{|l|c|} \hline
Uncertainty                         & $\Delta B/B$ \%           \\ 
Tracking efficiency                 &  5.0            \\ 
Luminosity                          &  3.6            \\ 
Kaon-id efficiency                  &  3.0            \\
Track resolution                    &  3.0            \\ 
Energy Resolution                   &  2.5            \\
Background shape                    &  2.3            \\ 
Monte Carlo Statistics              &  1.9            \\ 
Calorimeter energy scale            &  1.0            \\ 
Calorimeter efficiency              &  1.0            \\ 
\piz\/$\eta$ veto                   &  1.0            \\ 
Merged \piz\ modeling               &  1.0            \\ \hline
Total                               &  8.6            \\ \hline
\end{tabular}
\end{center}
\label{tab:systematic}
\end{table}
The efficiency for the selection of $\bkog$ candidates is $(15.6\pm 0.3)\%$.
The  branching fraction is measured to be  \brresult\  consistent both with 
previous measurements~\cite{CLEO} and with the standard model expectations
~\cite{theory}. The total systematic error of $8.6\%$ is a quadratic sum  
of several uncorrelated components given in Table~\ref{tab:systematic}~\cite{ichepbkg}.

\section{Search for 
$B^+ \rightarrow K^+ l^+ l^-$,
$B^0 \rightarrow \Kstarz \ l^+ l^-$}
We search in both the electron
and muon channels using a subset of the data corresponding to 
$3.7 \pm\ 0.1 \times\ 10^6$ \BB\ decays. 
The main goal of our study is to test the performance of 
a ``blind'' analysis in which 
the event selection is optimized without use of the signal
or sideband regions in the data. The dominant backgrounds 
come from random leptons and kaons in \BB\ and continuum
processes, and from $B\to \jpsi K^{(*)}$ or $B\to \psitwos K^{(*)}$ with 
$\jpsi$ or $\psitwos \to\ell^+\ell^-$. 
The $B$ candidates are reconstructed from $K^{*0}$, electron and muon candidates. The $K^{*0}$ is reconstructed in the $K^+\pi^-$ final state as above. 
Electron candidates are identified using the ratio of the deposited 
calorimeter energy to the associated charged track momentum. Muons
are identified by their depth of penetration into the muon detector.
Continuum backgrounds are suppressed using event shape variables~\cite{ichepkll}. The backgrounds from $B\to \jpsi K^{(*)}$ and $B\to \psitwos K^{(*)}$ are
eliminated by cutting in the   $\de$ vs. $M_{\ell^+\ell^-}$ plane ~\cite{ichepkll}.
Figure~\ref{fig:datambkll} shows the $\mes$ distributions for the 
four modes. 
\begin{figure}
\epsfxsize180pt
\figurebox{170pt}{160pt}{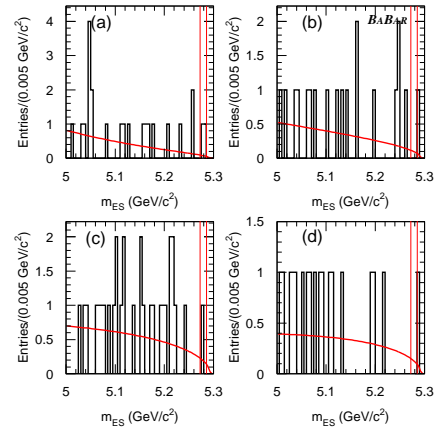}
\caption{$m_{ES}$ for data after all other event selection criteria are applied: (a) $\Bu\to \Kp e^+e^-$, (b) $\Bu\to \Kp \mu^+\mu^-$, (c) $\Bz\to \Kstarz e^+e^-$, and (d) $\Bz\to \Kstarz \mu^+\mu^-$.  The shape of the fit (the ARGUS function) is obtained from the large statistics sample of fast parameterized Monte Carlo events.  The lines indicate the signal region.}
\label{fig:datambkll}
\end{figure}
\begin{table}[hbtp]
\caption{Signal efficiencies, the number of
observed events, the number of estimated background events, and upper limits
on the branching fractions.}
\begin{center}
\begin{tabular}{|l|l|l|l|l|} 
\hline
Mode                     & Eff. & Obs.     & Bkg.  & $\BR/10^{-6}$ \\
                         & (\%) & evts     & est.  &  (90\% \\
                         &      &          &   &  C.L.)\\\hline
$K^+ e^+ e^-$             &  13.1 & 2 & 0.20  & $<$ 12.5  \\ 
$K^+ \mu^+ \mu^-$         &  8.6  & 0 & 0.25  & $<$ 8.3  \\  
$\Kstarz e^+ e^-$         &  7.7  & 1 & 0.50  & $<$ 24.1  \\
$\Kstarz \mu^+ \mu^-$     &  4.5  & 0 & 0.33 & $<$ 24.5  \\
\hline
\end{tabular}
\end{center}
\label{tab:resultskll}
\end{table}
No evidence of a signal is observed and table~\ref{tab:resultskll}
gives the derived limits on these processes. The derivation of the limits
takes into account the systematic uncertainties given in table~\ref{tab:syserrkll}~\cite{ichepkll}.
\begin{table}[hbtb]
\caption{Summary of the systematic uncertainties given as a percentage error on the
branching fraction}
\begin{center} 
\begin{tabular}{|l|c|c|c|c|} 
\hline
&\multicolumn{4}{c|}{$(\Delta \BR/\BR)$ (\%)} \\ \cline{2-5}
&${Kee}$ & ${K\mu\mu}$ & ${K^{*}ee}$ & ${K^{*}\mu\mu}$ \\ \hline 
Track eff.         & 7.5  &  7.5  & 10.0 & 10.0     \\
Lepton-id       & 4.0  &  5.0  &  4.0 &  5.0     \\
Kaon-id         & 3.0  &  3.0  &  3.0 &  3.0     \\
$\Delta E$  eff.
                            & 2.0  &  2.5  &  3.3 &  1.5  \\
$m_{ES}$  eff.        & 3.0  &  3.0  &  3.0 &  3.0  \\
Vertex  eff.       & 3.0  &  3.0  &  4.0 &  4.0  \\
Fisher  eff.       & 3.0  &  3.0  &  3.0 &  3.0  \\
Luminosity               & 3.6  &  3.6  &  3.6 &  3.6  \\      
MC stat.        & 3.4  &  4.0  &  4.3 &  6.1  \\ 
\hline
Total       & 11.7 & 12.3  & 14.2 & 14.8  \\ 
 \hline
\hline
\end{tabular}
\label{tab:syserrkll}
\end{center}
\end{table}

\end{document}